\documentclass{article}

\usepackage{url}
\usepackage{wrapfig}

\usepackage{graphicx}
\usepackage{tabularx}
\usepackage{makecell}
\usepackage{caption}
\usepackage{floatrow}
\usepackage{url}

\usepackage{listings}
\usepackage{color}

\definecolor{dkgreen}{rgb}{0,0.6,0}
\definecolor{gray}{rgb}{0.5,0.5,0.5}
\definecolor{mauve}{rgb}{0.58,0,0.82}

\graphicspath{flow/}
\graphicspath{KS/}
\graphicspath{mouse/}
\graphicspath{interface/}
\graphicspath{keystroke/}

\title{An experimental platform for gathering user behavioural data via browser APIs}
\date{\today}
\author{Zhaoyi Fan}

\begin{document}
\maketitle
\abstract 
Websites are capable of learning a wide range of information about the platform on which a browser is executing. One major source of such information is the set of standardised Application Programming Interfaces (APIs) provided within the browser, which can be accessed by JavaScript downloaded by a website; this information can then either be used by the JavaScript or sent back to the originating site. As has been widely discussed, much of this information can threaten user privacy. The main purpose of this paper is to document a publicly available platform designed to enable further investigation of one class of such threats, namely those based on analysing user behavioural data. The platform has two main components: a Chrome extension that gathers user keystroke and mouse data via browser APIs, and server software that collects and stores this data for subsequent experimentation.
\section{Introduction}
The JavaScript language is widely used by websites to enable dynamic behaviour. In the context of a web browser, JavaScript code is downloaded along with the web page. This JavaScript is executed automatically by the browser, and can access a variety of information via a set of browser-provided Application Program Interfaces (APIs). \par 
In general, APIs are constructs made available in programming languagues to allow developers to create complex functionality more easily. Browser-supported APIs, which can be accessed by website-originated JavaScript, provide website developers with more efficient ways to accomplish their goals and provide browser users with a better experience. \par 
In a modern browser, Client-side JavaScript can access a range of APIs \cite{WebAPI}. In general, browser APIs allow developers to create web pages that incorporate information from the user's environment \cite{W3CAPI}. In HTML5, a wide variety of information can be accessed via browser APIs, such as audio, application cache, canvas, fullscreen, geolocation, local storage, notifications, video and web database \cite{chrome}. However, browser APIs can also be used by websites to learn potentially privacy-sensitive information about the browser, the platform on which the browser is executing, and even the user of the browser. For example, platforms can be tracked across multiple web visits using browser fingerprinting techniques \cite{eckersley2010unique}, it may be possible to learn secret user data such as PINs \cite{simon2013pin} and the user location \cite{Nasser2017}, and it may even be possible to learn about the human user. \par 
The possibility of behavioural monitoring using browser APIs motivates the work described in this paper, namely the development of a platform to enable the collection of behavioural data. The platform has been designed to support experiments which aim to determine the degree to which individual users can be identified via browser APIs. When a web user is browsing a website, the standard Document Object Model (DOM) enables executing JavaScript to access information about keyboard and mouse events. These keyboard and mouse events can potentially be used to identify individual users using techniques developed for biometric identification and authentication \cite{Bailey2014User}. The platform we describe in the remainder of this paper has been designed to support experiments aimed at understanding how effective such identification can be. However, the platform is not restricted to this application, and is being made available as a potential tool for further research on browser security and privacy. \par 
The platform involves a Chrome extension that has been implemented to collect keystroke and mouse data from browser users. In addition, server functionality has been developed to receive and store this data. \par 
The remainder of the paper is structured as follows. Section 2 introduces the browser APIs used by the extension. Section 3 then addresses the implementation of the extension. Section 4 briefly describes the form of the data extracted by the extension. Section 5 introduces an approach to extracting usable data from the collected raw data. Section 6 describes possible ways in which the current platform might be extended, and Section 7 contains concluding remarks.\par
The experimental platform is made freely available for use by other researchers. It can be found at [\url{https://github.com/fanzhaoyi/DataCollector}]. Please send any comments (including bug reports) to: [\url{zhaoyi.fan.2016@live.rhul.ac.uk}]
\section{Background}
\subsection{Keyboard and mouse events}
We first introduce the browser API calls used by the experimental platform to gather information about user behaviour. Specifically we describe APIs which provide information about user interactions with keyboards and pointing devices (e.g. a mouse). \par 
The most commonly discussed browser API is probably the Document Object Model API which can create, remove and change HTML and CSS. The Document interface provides a representation of a web page loaded in the browser, and serves as an entry point into the web page content in the form of the DOM tree. The Event interface, one of the interfaces provided by DOM, provides the keyboard event and mouse event APIs. \par 
The KeyboardEvent API  provides information regarding user interactions with the keyboard. Each event is triggered by an action on a key, and can have an event type of \textit{keydown}, \textit{keypress} or \textit{keyup}. Table \ref{table:keyboard event} summarises the API calls for KeyboardEvent. \par 

\begin{table}[ht]
	\center
	\begin{tabular}{p{3.6cm}|p{5cm}}
		\hline
		\textbf{API call} & \textbf{Properties} \\
		\hline
		\textit{KeyboardEvent.code} & Returns a string with the code value of the key represented by the event.\\
		\textit{KeyboardEvent.key} & Returns the key value of the key represented by the event.\\
		\textit{keydown} & Triggered when the key is depressed.\\
		\textit{keyup} & Triggered when the key is released.\\
		\hline
	\end{tabular}
	\caption{API calls for keyboard events}
	\label{table:keyboard event}
\end{table}
The MouseEvent API provides information on events that occur when a user interacts with a pointing device, such as a mouse. It can give information regarding events such as moving a mouse, clicking a button, etc. Table \ref{table:mouse event} gives the API calls for MouseEvent. \par 
\begin{table}[ht]
	\center
	\begin{tabular}{p{2cm}|p{7cm}}
		\hline
		\textbf{API call} & \textbf{Properties} \\
		\hline
		\textit{click} & Triggered when the mouse button is pressed and released on a single element.\\
		\textit{mousemove} & Triggered when the pointer is moving while it is over an element.\\
		\textit{mousedown} & Triggered when the mouse button is pressed on an element.\\
		\textit{mouseup} & Triggered when the mouse button is released on an element.\\
		\textit{wheel} & Triggered when the mouse wheel is crolling over an element.\\
		\textit{pageX} & Returns the X coordinate of the mouse pointer relative to the whole document.\\
		\textit{pageY} & Returns the Y coordinate of the mouse pointer relative to the whole document.\\
		\textit{screenX} & Returns the X coordinate of the mouse pointer in global(screen) coordinates.\\
		\textit{screenY} & Returns the Y coordinate of the mouse pointer in global(screen) coordinates.\\
		\textit{clientX} & Returns the X coordinate of the mouse pointer in the applications’ client area. It changes when the page is scrolled.\\
		\textit{clientY} & Returns the Y coordinate of the mouse pointer in the applications’ client area. It changes when the page is scrolled.\\
		\hline
	\end{tabular}
	\caption{API calls for mouse events}
	\label{table:mouse event}
\end{table}
\subsection{Chrome extensions}
Chrome extensions are programs that can be used to customize the browsing experience. They enable users to tailor Chrome functionality and behaviour to individual needs or preferences. They are built on web technologies such as HTML, JavaScript, and CSS. By using the browser-supported APIs, an extension can collect user data and send it to a remote server. \par 
Extensions are made up of a range of possible components, which can include background scripts, content scripts, a manifest file and options pages. We next briefly review these component types. \par
The manifest file provides information about the extension. Typically the file \textit{manifest.json} contains general information about the extension (name, version, etc.), its permissions (download, urls, etc.) and settings for option pages. \par
Background scripts can work with or without a background HTML file. A background page is loaded when it is needed and unloaded when it becomes idle. It can communicate with content scripts and popup pages by sending messages and listening for an event. \par 
By using the DOM, content scripts can read details of web pages visited by the browser, make changes to them and pass information to other script pages. Working in an isolated environment, a content script is able to change the JavaScript environment of a loaded page without conflicting with the page or other scripts. It can be regarded as a part of the loaded page; however, a content script is not permitted to access any variables or functions created by the web page. The \textit{Cross-Origin XMLHttpRequest} policy prevents a content script from directly sending a \textit{XMLHttpRequest} to a remote server. Instead, message passing can be used to pass messages from a content script to the background script. \par 
\section{The experimental platform}
\begin{figure}
	\centering
	\includegraphics[scale=0.5]{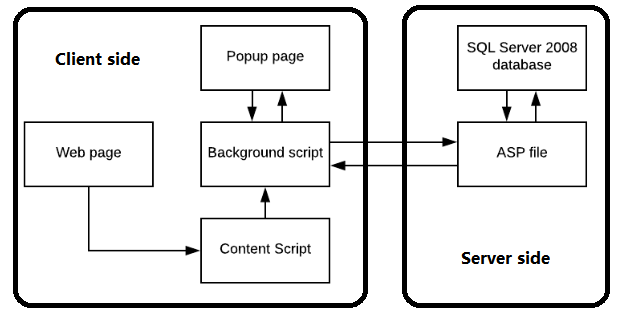}
	\caption{The structure of the experimental platform}
	\label{figure:flow}
\end{figure}

A Chrome extension was developed to collect user keystroke and mouse movement data. In addition, a server script was written to receive and store the data.  We next describe their development and operation. The structure of the experimental platform is shown in Figure \ref{figure:flow}.
\subsection{Development environment}
All software development was performed on a Windows 10 system, with details as follows. 
\begin{itemize}
	\item \textbf{Programming tool}: Sublime Text (Version 3.1.1, Build 3176).
	\item \textbf{Programming languages:} HTML and JavaScript (client-side), ASP (server-side), Python (data processing).
	\item \textbf{Google Chrome version:} 68.0.3440.106 (Official Build).
\end{itemize}
To set up the server, an Alibaba Cloud Web hosting service was used, which supports the ASP language and has a 1GB SQL database. \par 

\subsection{The Chrome extension} 
The extension is made of a manifest.json file, a popup HTML page, a popup.js script file, a background.js script file and a content.js script file. The popup page is shown when a user clicks the extension. The popup page allows the user to register and then log in. After the user has logged in, the content script starts monitoring visited webpages. Keyboard event APIs and mouse event APIs are used to monitor the user{'}s keyboard and mouse activities. Since content scripts cannot directly communicate with a remote server, each time the keyboard or mouse event is triggered, the sendMessage method is called to pass the message from the content script to the background script. When the background script receives the data, an XMLHttpRequest is used to interact with the server. \par 
As shown in Figure \ref{figure:flow}, at the client the content script extracts keystroke and mouse data using browser APIs and sends the data to the background script via message passing. The popup page communicates with the background script and informs the user about the login status. 
The permissions are set in \textit{manifest.json}. The extension can access any URL and use cookies. \par 
Content scripts can read the DOM of a web page. The content script uses six event listeners: \textit{mousemove, mousedown, mouseup, wheel, keydown} and \textit{keyup}. Every time such an event is triggered, a timestamp is created and the corresponding data is captured. The data shown below is then sent to the background script. \par 
\begin{lstlisting}
    chrome.runtime.sendMessage({
        type: actiontype,
        data: actiondata
        },function(response){
    })
\end{lstlisting}

Action type has two possible values: keystroke and mouse. For a single keystroke, the data is regarded as an object with eight elements: keystroke timestamps, key value, and the usage of functional keys (Ctrl key, Alt key, Shift key and CapsLock key), as shown in Figure \ref{figure:ks raw data}. When a key is depressed, an object of type keystroke with eight elements is created. When the key is released, the uptime is updated and the data is sent to the background script. \par 
For mouse actions the data consists of one of eight mouse action types (mouse move, left button down, left button up, right button down, right button up, wheel rolls, wheel down and wheel up), the coordinates of the mouse (X and Y coordinates) and the timestamp, as shown in Figure \ref{figure:mouse raw data}.  \par 
The background script receives the request message from content script and popup script:
\begin{lstlisting}
 chrome.runtime.onMessage.addListener(
     function(request, sender, sendResponse){
         if (request.type == action type ){
             SendToServer({
                 data: request.data,
                 type: request.type,
             })
         }
     }
 )
\end{lstlisting}
The main function of the background script is communicating with the other scripts and the server. When the background script receives a message from another script, an \textit{XMLHttpRequest} is established to send the data to remote server: 
\begin{lstlisting}
  xmlhttp.open("GET", URL+data, TRUE)
  xmlhttp.send()
\end{lstlisting} 
The popup page allows the user to register and login. When the user logs in, a message is passed to the server via the background page. \par 
\begin{figure}
	\centering
	\includegraphics[scale=0.5]{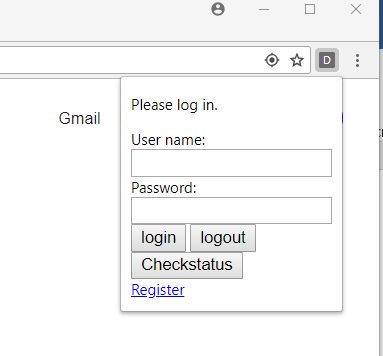}
	\caption{Popup page user interface}
	\label{figure:interface}
\end{figure}
As the extension can record every keystroke and mouse movement, some personal data might be gathered, including passwords for other websites and other sensitive data. Thus if a user has any concerns that the data they are typing is possibly sensitive, the user is recommended to log out of the extension by clicking the logout button; the user can log in again when the sensitive interactions have been completed. \par 

\subsection{The server} 
At the server, ASP files are used to interact with the client and database. The data is stored in a SQL Server 2008 database. To prevent  SQL injection attacks, a regular expression check is used when data is received. To connect with the SQL server database, \textit{ADODB.CONNECTION} is used in the ASP file:
\begin{lstlisting}
  Set Conn=server.CreateObject("ADODB.CONNECTION")
  StrConn=" Provider=SQLOLEDB;Data Source= domain;User ID= userID;Password= pwd;Initial Catalog= databaseName"
  Conn.open StrConn
\end{lstlisting}

When a user registers on the popup page, their registration information is sent to the server. To distinguish data sent from different devices, the server creates a cookie for each user:
\begin{lstlisting}
  response.cookies("uname") = username
\end{lstlisting}

When a user logs out, the server deletes the cookie. The server maintains two databases for each user, one to store keystroke data and the other to store mouse use data. \par 

\section{Data collection}
When the user is logged in and the browser web page is active, the extension monitors every keyboard and mouse event performed by the user. 
\subsection{Raw keystroke data}
\begin{figure}
	\centering
	\includegraphics[scale=0.5]{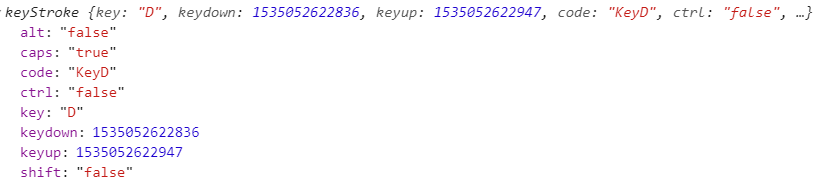}
	\caption{Raw keystroke data}
	\label{figure:ks raw data}
\end{figure}

Each keystroke is regarded as a single object containing the following elements: key code, key value, key down time, key up time, Ctrl, Alt, Shift, CapsLock. Figure \ref{figure:ks raw data} gives an example of the data generated if a user presses the key ‘D’. \textit{Keydown} and \textit{keyup} are timestamps, \textit{code} represents the physical key position on the keyboard, while \textit{key} is the key value that the user has typed. \textit{alt}, \textit{ctrl}, \textit{shift} and 
\textit{caps} indicate use of these four function keys. It should be noted that, to type an uppercase letter, a user can either use the Caps Lock key or hold down the Shift key. 

\subsection{Raw mouse data}
\begin{figure}
	\centering
	\includegraphics[scale=0.5]{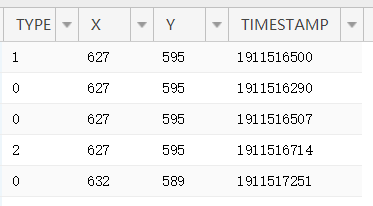}
	\caption{Raw mouse data}
	\label{figure:mouse raw data}
\end{figure}

A keyboard event only has one type of action: press. However, a mouse event can be triggered by a number of different types of action, such as mouse movement, left button click, right button click and wheel scrolling. Thus each mouse event is regarded as a single object containing the following elements: type of event action, X coordinate of the mouse, Y coordinate of the mouse, and a timestamp, as shown in Figure \ref{figure:mouse raw data}. \par 

\section{Data processing}
Depending on the use to be made of the collected data, it is first necessary to process it to extract the features to be used. As examples, we next describe possible approaches to such processing, used in experiments that were performed with the aid of the platform.
\subsection{Feature extraction for keyboard data}
\begin{figure}
	\centering
	\includegraphics[scale=0.5]{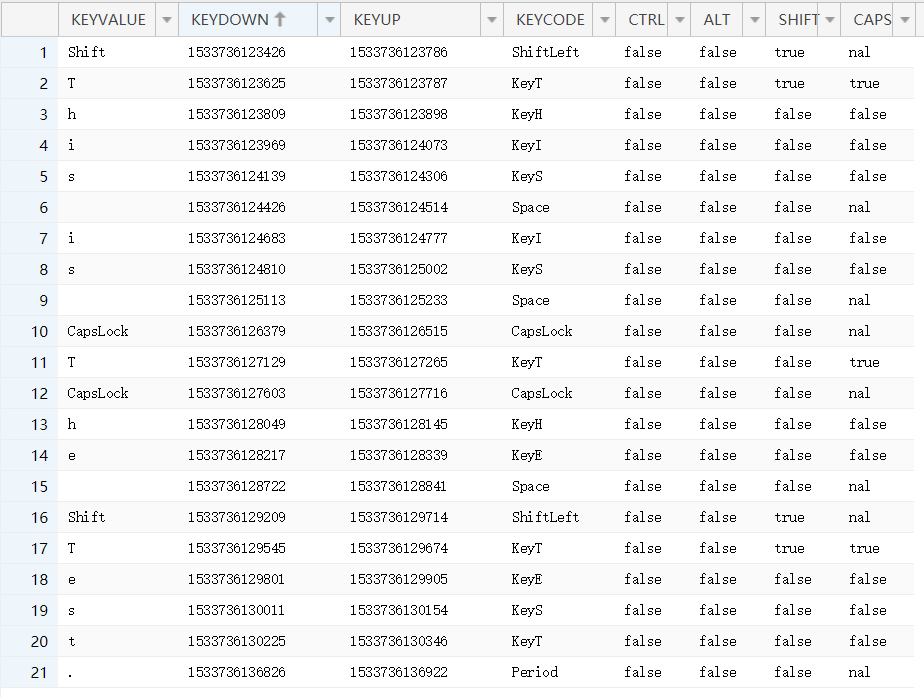}
	\caption{Keystroke raw data in the database}
	\label{figure:20ks}
\end{figure}

Figure \ref{figure:20ks} gives an example of raw keystroke data as stored in the server database. This data was processed in the following way to enable it to be used for possible user identification based on how letter bigrams are typed. The raw data is first divided into a sequence of sets using the values of the keydown timestamp. \par 
The sequence of keystrokes is divided into discrete sets using two rules: \par 

\begin{itemize}
	\item if the time interval between two consecutive keystrokes is greater than one second, or
	\item if the key depressed is the space key,
\end{itemize}

then a new set is started. Keystrokes corresponding to the space key and to function keys are removed from the data set. Using these rules, the data shown in Figure \ref{figure:20ks} will yield the four sets (words) ‘This Is The Text’.

The sequence of letters in each set was then used to derive detailed timing information for each pair of consecutive letters (or bigraph). For each such bigraph four timestamps were extracted, namely the key down and key up times for the two letters. The biograph timing information was then used to try to identify the user. Of course, this is only one approach to extracting usable data from the raw data, and is provided here for illustrative purposes only. \par 

\subsection{Feature extraction for mouse data}
For the purposes of user identification two types of data were extracted from pairs of consecutive mouse events: Euclidean distance between the locations and elapsed time between the events. This was then used to compute the mouse movement speeds between events.  The movement speed between a range of pairs of types of mouse event were then used to attempt to identify users.
 
\section{Possible future work}
The experimental platform is currently capable of gathering keyboard event and mouse event data. However, there is other behaviour-related (and hence privacy-sensitive) data that could be gathered by the platform. For example, motion sensor data for a mobile device is accessible via motion sensor APIs. Extensions to the platform are envisaged which would gather such data.

\section{Conclusions}
This document describes the functioning and use of an experimental platform designed to gather user-related data from a browser. The platform has been developed to support experiments aimed at understanding the degree to which individual users can be identified using browser-gathered behavioural data. It is hoped that the platform will be of value to other researchers performing similar experiments. \par 
Section 2 introduces the browser APIs used by the experimental platform. Section 3 provides details of the implementation. Section 4 describes the form of the data extracted by the extension. Section 5 illustrates possible approaches to extracting usable data from the collected data. Section 6 describes possible further development of the platform. \par

\bibliographystyle{plain}
\bibliography{re}
\end{document}